\documentclass[twocolumn,showpacs,preprintnumbers,amsmath,amssymb]{revtex4-1}
\usepackage{amsmath, amssymb}
\usepackage{natbib}
\usepackage{graphicx}
\usepackage{setspace}
\usepackage{gensymb}
\usepackage{dcolumn}
\usepackage{bm}
\usepackage[usenames,dvipsnames]{color}
\usepackage{xcolor}
\usepackage{bm}
\begin{document}
\def\bk{\bf k}
\newcommand{\be}{\begin{equation}}
\newcommand{\ee}{\end{equation}}
\newcommand{\beq}{\begin{eqnarray}}
\newcommand{\eeq}{\end{eqnarray}}

\title{Thermodynamic equivalence of two-dimensional \\imperfect attractive Fermi and repulsive Bose gases}

\author{Marek Napi\'{o}rkowski and Jaros{\l}aw Piasecki}
 
\affiliation{Institute of Theoretical Physics, Faculty of Physics \\ University of Warsaw, 
 Pasteura 5, 02-093 Warsaw, Poland}

\date{\today}
\begin{abstract}
We consider two-dimensional imperfect attractive Fermi and repulsive Bose gases consisting of spinless point particles whose total  interparticle interaction energy is represented by $a N^2/2 V$ with $a=-a_{F}\leq 0$ for fermions, and $a=a_{B}\geq 0$ for bosons. We show that in spite of the attraction the thermodynamics of $d=2$ imperfect Fermi gas remains well defined for $0 \leq a_{F}\leq a_{0}=h^2/2\pi m$, and is exactly the same as the one of the repulsive imperfect Bose gas with $a_{B}=a_{0}-a_{F}$. In particular, for $a_{F}=a_{0}$ one observes the thermodynamic equivalence of the attractive imperfect Fermi gas and the ideal Bose gas. 
\end{abstract}


\maketitle

\section{Introduction}

The fundamental and subtle problem of the existence of thermodynamics for interacting quantum gases has always been of utmost interest and has huge literature, see e.g.,  \cite{Lieb1,Lieb2,LS1} and references therein. In the simplest case, if the point quantum particles, whether fermions or bosons, repel each other there is no danger of a collapse  and  provided the thermodynamic limit is well defined the thermodynamic description exists. On the other hand, if the particles attract each other the situation becomes different. For bosons the answer is simple: the pure attraction between bosons rules out the existence of stable equilibrium and in this sense leads the collapse of the system. For fermions, however, the Pauli principle, which already for the ideal Fermi gas leads to the effective repulsion, can play a decisive role and counterbalance the interparticle attraction. The object of this study is to investigate to what extent the quantum statistics plays a constructive role and provides the existence of thermodynamics for attractive fermions. And, can one expect that the Fermi statistics will not only counterbalance the attraction but effectively lead to the thermodynamic behavior similar to that of a repulsive boson gas. A resemblant effect is known in the BCS-theory of superconductivity and recently has been intensely investigated in relation to the BCS-BEC crossover 
\cite{Zwerger1}. The accompanying more detailed question is how this possible balance between interaction and statistics depends on the dimensionality of the system  and on the strength of the  interaction. \\

In the following we give an answer to these questions by considering the model of the so-called imperfect, spinless quantum gases 
\cite{Huang1,Davies1,BP1983,Zag2001,Lewis_book,Berg_84,NP1,NJN1}. In the occupation number representation the Hamiltonian of the imperfect Fermi gas has the following form 
\be
\label{HPG}
H_{F,imp} = \sum_{\bk}\frac{\hbar^2 \bk^{2}}{2m} n_{\bk}  - \frac{a_{F} }{2}\frac{N^2}{V}  \quad, \,\,\, a_{F} \geq 0 \quad, 
\ee 
where $V$ denotes the volume of the system, $N = \sum\limits_{\bk}n_{\bk}$, $n_{\bk}$ is the occupation number of one-particle state with momentum $\hbar{\bf{k}}$, $n_{\bk}\,=\,0,1 $. The coupling constant $a_{F}$ measures the strength of the mean-field attractive potential energy 
$-a_{F}N^2/2V$.  Analogous expression holds for the Hamiltonian of the repulsive imperfect Bose gas with the coupling constant $a_{F}$ replaced by $-a_{B} <0$ and $n_{\bk}\,=\,0,1, \dots \infty$. The model of imperfect quantum gases has been succesfully applied to repulsive  bosons \cite{Huang1,Davies1,BP1983,Zag2001,Lewis_book,Berg_84,NP1,NJN1} where, inter alia, the Bose-Einstein condensation and the Casimir forces were discussed. In this paper we discuss both the attractive imperfect Fermi gas and the repulsive imperfect Bose gas and show their thermodynamic equivalence in two dimensions.  

\section{The imperfect Fermi gas}

We work in the grand canonical ensemble parametrized by temperature $T$, chemical potential $\mu$, volume $V$, and assume periodic boundary conditions. The evaluation of the grand canonical partition function 
\begin{eqnarray}
\label{part1}
\Xi_{F,imp}(T,V,\mu) = \nonumber \\ \sum_{N=0}^{\infty} \,\sum\limits_{\{n_{\bk}\}} {}^{'} \,\exp\left[-\beta \left(\sum_{\bk} n_{\bk} 
(\epsilon_{k}-\mu) - \frac{a_{F}}{2}\frac{N^2}{V} \right)\right],
\end{eqnarray}
where $\beta=1/k_{B}T$, is stymied by the presence of the factor $\exp{\left(\frac{\beta a_{F} N^2}{2V}\right)}$ which calls into question the very existence of 
$\Xi_{F,imp}(T,V,\mu)$. The ground state energy of $d$-dimensional ideal Fermi gas at fixed volume behaves as $E_{0} \sim N^{1+\frac{2}{d}}$ which hints at the possibility that for $d \leq2 $ the grand canonical partition function may exist. In order to resolve this issue we  use the identity
\beq
\label{tozs1}
\exp{\left(\frac{\beta a_{F}}{2 V}\, N^2 \right)} = \nonumber \\ \left(\frac{V}{2 \pi\beta a_{F}}\right)^{\frac{1}{2}}  
\int\limits_{-\infty}^{\infty} dq  \exp{\left(-\,\frac{V q^2}{2\beta a_{F}} -N q\right)}   
\eeq
and rewrite $\Xi(T,V,\mu)$ as 
\begin{eqnarray}
\label{part2}
\Xi_{F,imp}(T,V,\mu) = \hspace{1cm} \\  
\left(\frac{V}{2 \pi\beta a_{F}}\right)^{\frac{1}{2}} \int\limits_{-\infty}^{\infty} 
dq \exp{\left(-\frac{Vq^2}{2\beta a_{F}}\right)} \Xi_{F,id}\left(T,V,\mu-\frac{q}{\beta}\right) \nonumber ,  
\end{eqnarray}
where $\Xi_{F,id}(T,V,\mu)$ denotes the grand canonical partition function of an ideal Fermi gas. For large $V$  Eqs~(\ref{part2}) can be represented as 
\beq
\label{part4}
\Xi_{F,imp}(T,V,\mu)\,= \nonumber\\ \left(\frac{V}{2 \pi\beta a_{F}}\right)^{\frac{1}{2}} \int\limits_{-\infty}^{\infty} dq \exp{[- V \varphi_{F}(q;T,\mu)]}, 
\eeq
with 
\begin{widetext}
\beq
\label{part5}
\varphi_{F}(q;T,\mu)  =   \frac{1}{\lambda^d} \, \left[ \frac{\lambda^{d-2}}{2} \frac{a_{0}}{a_{F}} (q-\beta \mu )^2 - 
\frac{1}{\Gamma(\frac{d}{2}+1)} \int\limits_{0}^{\infty} dx \frac{x^{\frac{d}{2}}}{1+e^{x-q}} \right] , 
\eeq 
\end{widetext}
\noindent where $a_0 = h^2/2\pi m$ and $\lambda$ is the thermal de Broglie wavelength, $\beta a_{0} = \lambda^2$. Thus the problem of the existence and evaluation of the series in Eq.~(\ref{part1}) has been rephrased as the question of the existence of the integral in Eq.~(\ref{part4}). Once the conditions for its existence are settled the bulk thermodynamics can be extracted from it via the method of steepest descent. \\
The parameter $a_0$ turns out to play a prominent role in our analysis and for this reason its presence in Eq.~(\ref{part5}) is explicitly exposed. The existence of the integral in Eq.~(\ref{part4}) depends on the behavior of function $\varphi_{F}(q;T,\mu)$ for large $|q|$. When 
$q \rightarrow -\infty$ one has $\varphi_{F}(q;T,\mu)/q^2 \sim a_{0} / 2 a_{F} \lambda^2$ and this regime poses no problem. On the other hand  for $q \rightarrow \infty$ one has  
$\varphi_{F}(q;T,\mu)/q^2 \sim  - q^{\frac{d}{2}-1}/ \lambda^d \Gamma(\frac{d}{2}+2) \, + \, a_{0}/2 a_{F} \lambda^2 $. Thus for $d>2$ the grand canonical partition function ceases to exist (note that this observation refers to fermions with purely attractive interactions). For $d=2$, when $q \rightarrow \infty$ one finds  $\varphi_{F}(q;T,\mu)/q^2 \sim (a_0 - a_{F})/2 a_{F} \lambda^2$ which delimits the allowed values of the coupling constant $a_{F}$ to the range $0 \leq a_{F} \leq a_{0}$. 
For the boundary value $a_{F}=a_0$ one has $\varphi_{F}(q;T,\mu) \rightarrow -\frac{\beta \mu}{\lambda^2} q$ which means that in this particular case only negative values of chemical potential are allowed. This limitation will be shown to have a simple physical interpretation. Note that in two dimensions parameter $a_{0}$ has a straightforward interpretation: the ground state energy per particle 
$e_{0}= n (a_0 - a_F)/2$, where $n$ is the density, becomes negative for $a_{F} > a_{0}$.  

\section{Thermodynamics of the two-dimensional imperfect Fermi gas}

\begin{figure}
\begin{center}
\includegraphics[width=8cm]{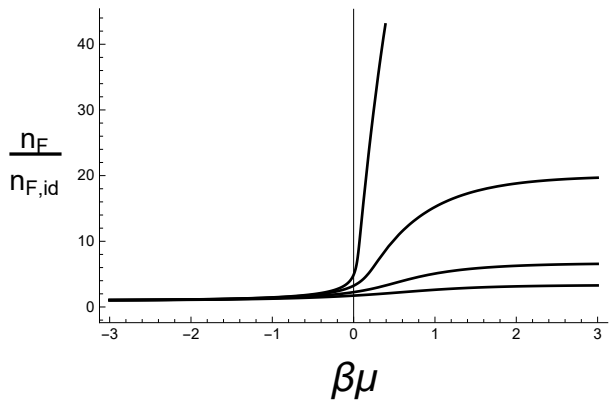}
\caption{\label{fig:density} The plot of the imperfect Fermi gas density $n_{F}$ normalized by the ideal Fermi gas density 
$n_{F,id}$ as a function of $\beta \mu$ for $a/a_{0}=0.7, 0.85, 0.95, 0.99$ (larger $a/a_{0}$ values correspond to larger densities).  }
\end{center}
\end{figure}

The thermodynamic limit of the grand canonical potential density 
$\lim\limits_{V \rightarrow \infty} \frac{\Omega_{F,imp}(T,V,\mu)}{V} = \omega_{F}(T,\mu) = - p_{F}(T,\mu)$ can be calculated using the method of steepest descent. 
It gives 
\beq
\label{varB1}
\omega_{F}(T,\mu) = k_{B}T \, \varphi_{F}(q_{0}(T,\mu);T,\mu) , 
\eeq  
where $q_{0}(T,\mu)$ minimizes $\varphi_{F}(q;T,\mu)$ and fulfills the equation
\beq
\label{q0}
q_0 = \beta \mu + \frac{a_{F}}{a_0} \ln{\left(1+ e^{q_{0}}\right)} .
\eeq
The number density $n_{F}(T,\mu) = - \left(\frac{\partial \omega_{F}}{\partial \mu}\right)_{T} $ is thus related to $q_0$ via 
\beq
\label{density}
q_0(T,\mu) = \beta (\mu + a_{F} n_{F}(T,\mu)) 
\eeq 
while the pressure $p_{F}(T,\mu)$ is  
\be
\label{eqstate1}
p_{F}(T,\mu) = - \,\frac{a_{F}}{2} n_{F}^2(T,\mu) + p_{F,id}(T, \mu + a_{F} \, n_{F}(T,\mu)),
\ee
where  $p_{F,id}(T,\mu)$ denotes the pressure of the ideal Fermi gas
\beq
\label{eqstateidB}
p_{F,id}(T,\mu) =  \frac{k_{B}T}{\lambda^2} \int\limits_{0}^{\infty} dx \, \ln\left(1+e^{\beta\mu-x}\right). 
\eeq
The density $n_{F}(T,\mu)$ is obtained by solving Eq.~(\ref{q0}) for $q_{0}(T,\mu)$ and inserting the result into Eq.~(\ref{density}) 
\beq
\label{denF1}
n_{F}(T,\mu)= \frac{1}{\lambda^2} \ln{\left(1+e^{q_{0}(T,\mu)}\right)},  
\eeq
see Fig.~\ref{fig:density}. 
The equation of state, $p_{F}(T,n)$ takes the form 
\beq
\label{eqstate2}
p_{F}(T,n) = -\,\frac{a_{F}}{2} n^2 + p_{F,id}(T,n),
\eeq
where
\be
\label{pFid1}
p_{F,id}(T,n) = - \frac{k_{B}T}{\lambda^2}  \sum\limits_{r=1}^{\infty} \frac{\left(1-e^{n \lambda^2}\right)^r}{r^2} .
\ee 

The coefficient of isothermal compressibility $\chi_{T}(T,n)=\frac{1}{n} \left(\frac{\partial n}{\partial p}\right)_{T}$ 
of the imperfect Fermi gas has the following form
\beq
\label{kap1}
\chi_{T}(T,n) = \left[  n^2 (a_0 - a_{F})  + \frac{n^2 a_0}{e^{n\lambda^2}-1} \right]^{-1}. 
\eeq 
For $a_{F} < a_0$ it remains finite. However, for $a_{F}=a_{0}$ the compressibility becomes infinitely large for 
$n \rightarrow \infty$ at fixed $T$, or equivalently for $\mu \rightarrow 0^{-}$ at fixed $T$ (in this limit $n_{F}\lambda^2 = -\ln{(-\beta\mu)}$).  

\section{Thermodynamic equivalence of two-dimensional imperfect Fermi and Bose gases}
 
For the special case $a_{F}=a_0$, it follows from Eqs~(\ref{q0}-\ref{eqstateidB}) that 
\be
\label{den10}
n_{F}(T,\mu) = - \frac{1}{\lambda^2} \ln{\left(1-e^{\beta \mu}\right)}
\ee
and  
\be
\label{eq10}
p_{F}(T,n) = - \frac{k_{B}T}{\lambda^2} \left[ \frac{\left(n\lambda^2 \right)^2}{2} + 
\sum\limits_{r=1}^{\infty} \frac{(-1)^r}{r^2} \left(e^{n\lambda^2} -1\right)^r \right].
\ee  

One notes that remarkable identities follow from the above formulas.  They relate the imperfect Fermi gas at $a_{F}=a_0$  
and the ideal Bose gas \cite{ZKU} (both defined for $\mu < 0$) for which 
\be
\label{denB1}
n_{B,id}(T,\mu) = - \frac{1}{\lambda^2} \ln{\left(1-e^{\beta\mu}\right)}
\ee
and 
\beq
\label{presB1}
p_{B,id}(T,n) = \frac{k_B T}{\lambda^2} \sum\limits_{r=1}^{\infty} \frac{e^{\beta\mu(T,n) r}}{r^2} = \nonumber \\
 \frac{k_BT}{\lambda^2} \sum\limits_{r=1}^{\infty} \frac{\left(1-e^{-n\lambda^2}\right)^r}{r^2}. 
\eeq
It follows from Eqs~(\ref{den10}) and (\ref{denB1}) that for $a=a_{0}$ 
\beq
\label{idn1}
n_{F}(T,\mu) = n_{B,id}(T,\mu) 
\eeq 
while from Eqs~(\ref{eq10}) and (\ref{presB1}) one finds  
\beq
\label{idp1}
p_{F}(T,n) = p_{B,id}(T,n). 
\eeq 
The above equality can be checked using the integral representation of function $p_{B,id}(T,\mu)$ in Eq.~(\ref{presB1}) and  
$p_{F,id}(T,\mu)$ in Eq.~(\ref{pFid1}). A straightforward calculation leads to 
\beq
\label{ppFB}
\left.\frac{\partial}{\partial n}\right|_{T} \left[ p_{F}(T,n) - p_{B,id}(T,n) \right] =0
\eeq 
from which Eq.~(\ref{idp1}) follows because $\left[p_{F}(T,n) - p_{B,id}(T,n)\right]|_{n=0} = 0$.  \\

Thus according to Eqs~(\ref{idn1}) and (\ref{idp1}) the two-dimensional imperfect Fermi gas at $a=a_0$ and the ideal Bose gas, both in the same state characterized by arbitrary $T$ and $\mu<0$ have the same densities $n_{F}(T,\mu)=n_{B,id}(T,\mu)$ and pressures 
$p_{F}(T,n)=p_{B,id}(T,n)$.  \\ 

In order to put this remarkable equivalence into a broader perspective \cite{Rand1,Drechsler1,Bertaina2011,Makhalov2014,Bauer1,Murthy2015} 
we analyze the two-dimensional imperfect, spinless Bose gas with repulsive interactions characterized by the coupling constant $a_{B} > 0$
\be
\label{BHMF}
H_{B,imp}  = \sum_{\bk}\frac{\hbar^2 \bk^{2}}{2m} n_{\bk} + \frac{a_{B} }{2}\frac{N^2}{V} . 
\ee
As far as the existence of thermodynamics is concerned there is no upper bound on the coupling constant $a_B$. The repulsive imperfect 
Bose gas has been intensely discussed in the literature, see e.g., \cite{Huang1,Davies1,BP1983,Zag2001,Lewis_book,Berg_84,NP1,NJN1,Diehl1}. 
For $d>2$ it shows Bose-Einstein condensation taking place for $\mu > a_{B} \zeta(d/2) \lambda^{-d}$ with the critical indices belonging to the mean-spherical model universality class. The formalism used in \cite{NP1,NJN1} to evaluate the partition function is analogous to the one employed here with the identity in Eq.~(\ref{tozs1}) taking the role of the Hubbard-Stratonovich transformation \cite{Hubbard1} used in \cite{NP1,NJN1}. Thus we only quote the relevant formulas. Analogously to the Fermi case, the grand canonical free energy density 
$\omega_{B}(T,\mu)$ can be  obtained as the minimum of 
\be
\label{part6}
\varphi_{B}(s;T,\mu)\,=\, -\frac{1}{\lambda^2}\,\left[\frac{a_{0}}{2a_{B}} \left(s-\beta\mu \right)^2 + \,g_{2}\left(\exp(s)\right)\right] 
\ee
with respect to variable $s$, where $g_{2}(z)$ is the Bose function, for details see \cite{NP1,NJN1}. The thermodynamics follows from the following set of equations 
\beq 
\label{impB1}
\omega_{B}(T,\mu) = -p_{B}(T,\mu) = k_{B}T \varphi_{B}(s_0(T,\mu);T,\mu)
\eeq
where the parameter $s_0(T,\mu)$ minimizing the function $\varphi_{B}(s;T,\mu)$ solves the equation
\beq
\label{impB2}
s_0 - \beta\mu  = \frac{a_B}{a_{0}} \ln{\left(1-e^{s_0}\right)} .
\eeq
The density $n_{B}(T,\mu)$ is related to $s_{0}(T,\mu)$ via
\beq
\label{s0}
s_{0}(T,\mu)=\beta(\mu-a_{B}n_{B}(T,\mu))
\eeq
and can be rewritten as
\beq
\label{impB3}
n_{B}(T,\mu) \lambda^2 = - \ln{\left(1-e^{s_0(T,\mu)}\right)} 
\eeq
which gives $p_{B}(T,n)$ in the following form 
\beq
\label{impB4}
p_{B}(T,n) =  \frac{a_{B}}{2} n^2 + \frac{k_{B}T}{\lambda^2} g_{2}\left(1-e^{-n\lambda^2}\right) . 
\eeq 
\\

The above set of equation Eqs~(\ref{impB1}-\ref{impB4}) can be confronted with the corresponding set Eqs~(\ref{varB1}-\ref{eqstateidB}) for the attractive imperfect Fermi gas. In order to find the relation between the imperfect attractive Fermi and the imperfect repulsive Bose gas 
\cite{Zwerger1,Rand1,Drechsler1,Bertaina2011,Makhalov2014,Bauer1,Murthy2015,AndersonDrut2015} we rewrite Eq.~(\ref{denF1}) in the following form
\beq
n_{F} \lambda^2 = - \ln{\left(1-e^{\zeta_0}\right)}
\eeq
where the parameter $\zeta_{0}$ is defined via 
\beq
\label{zeta1}
\zeta_0=-\ln{\left(1+e^{-q_0}\right)} .
\eeq
It follows from Eqs~(\ref{zeta1}) and (\ref{q0}) that 
\beq
\zeta_{0} - \beta\mu  = \left(1-\frac{a_{F}}{a_{0}}\right) \ln{\left(1- e^{\zeta_{0}}\right)}  
\eeq 
and thus the parameter $\zeta_{0}$ fulfills the same equation as parameter $s_{0}$, see Eq.~(\ref{impB2}), provided 
\beq
\label{sym1}
a_B = a_0 - a_F .
\eeq
In other words, if the above relation holds then the fermion and boson densities are identical:  
 $n_{F}(T,\mu) = n_{B}(T,\mu)$. 
Similarly, it follows from Eqs~(\ref{eqstate1}),(\ref{eqstate2}),(\ref{presB1}),(\ref{impB4}) that 
\beq
\label{presBF1}
p_{F}(T,n) = \frac{n^2}{2} (a_0-a_F) + p_{B,id}(T,n) = \nonumber \\ \frac{n^2}{2} (a_0-a_F-a_B) + p_{B}(T,n) 
\eeq
and thus when the relation (\ref{sym1}) is fulfilled one has $p_{F}(T,n) = p_{B}(T,n)$. Thus for each value of the coupling constant 
$a_F \leq a_{0}$ there exists a  repulsive imperfect Bose gas characterized by the coupling constant $a_B = a_0 - a_F$; it has the same density and pressure as the Fermi gas. In particular, when $a_F=a_0$ one has $a_{B}=0$ and the previously proven result $n_{F}(T,n) = n_{B,id}(T,n)$ and $p_{F}(T,n) = p_{B,id}(T,n)$, see 
Eqs~({\ref{idn1},\ref{idp1}) is recovered. In this case the attractive mean field suppresses the effect of the Fermi statistics and makes the pressure of the Fermi gas approach zero for $T \rightarrow 0$ at any density $n$. On the other hand, for $a_{F}=0$ one has 
$p_{F,id}(T,n)= a_{0}n^2/2 + p_{B,id}(T,n)$. These are special cases of the general equivalence condition in Eq.~(\ref{sym1}). \\

To summarize, we have shown that thermodynamics exists for the two-dimensional attractive imperfect Fermi gas provided the coupling constant 
$a_{F}$ measuring the strength of the attractive energy is not too large, $0 \leq a_{F} \leq a_{0}$. No thermodynamics exists for three and higher dimensional imperfect attractive Fermi systems. We showed that in two dimensions the density and pressure of the attractive Fermi gas are the same as those of the repulsive imperfect Bose gas provided the coupling constant $a_{B}$ measuring the strength of repulsion fulfills the relation $a_{B}=a_{0}-a_{F}$. Needless to say, the above derived equivalence of thermodynamics of imperfect attractive Fermi and repulsive Bose gases is restricted to two-dimensions where the imperfect Bose gas does not suffer the Bose-Einstein condensation.

\begin{acknowledgements} 
M.N. acknowledges the support from National Science Center, Poland via grant 2014/15/B/ST3/02212.
\end{acknowledgements}

\end{document}